
%
%
%
\def\gsim{\mathrel{\scriptstyle{\buildrel > \over \sim}}}
\def\lsim{\mathrel{\scriptstyle{\buildrel < \over \sim}}}
\magnification 1200
\baselineskip=17pt

\centerline{\bf BEREZINSKII-KOSTERLITZ-THOULESS TRANSITION}
\bigskip
\centerline{{\bf IN SPIN-CHARGE SEPARATED SUPERCONDUCTOR}\footnote {*}
{Accepted for publication in Physical Review B.}}
\vskip 50pt
\centerline{J. P. Rodriguez}
\medskip
\centerline{Dept. of Physics and Astronomy,
California State University, Los Angeles, CA 90032.
}
\vskip 30pt
\centerline  {\bf  Abstract}
\vskip 8pt\noindent
A  model
for spin-charge separated superconductivity
in two dimensions is introduced where  the phases of the spinon
and holon order parameters
couple gauge-invariantly to a statistical gauge-field
representing chiral spin-fluctuations.
The model is analyzed in  the continuum
limit and in the low-temperature limit.
In both cases we find that physical electronic
phase correlations show a
superconducting-normal phase transition of the
Berezinskii-Kosterlitz-Thouless type,
while statistical gauge-field excitations
are found to be strictly gapless.  It is
argued that the former transition is in
the same universality class as that of the XY model.
We thus predict a universal jump
in the superfluid density at
this transition. The normal-to-superconductor
phase boundary for this model is also obtained
as a function of carrier density, where we
find that its shape compares favorably with
that of the experimentally observed phase
diagram for the  oxide superconductors.

\bigskip
\noindent
PACS Indices:  74.20.Mn, 11.15.Ha, 71.27.+a
\vfill\eject
\centerline{\bf I. Introduction}
\bigskip
Although the phenomenon of
high-temperature superconductivity$^1$ has sparked a
tremendous amount of theoretical activity, a viable theory
remains to be uncovered.  It is, however,
generally agreed upon that
strong electron-electron interactions
must play an
important role in the charge dynamics of
the Copper-Oxygen planes that are common
to these systems.$^2$
Among the various attacks on the theoretical problem of strongly
interacting electrons in
two dimension, those based on the hypothesis
of spin-charge separation
show promise at the phenomenological level.$^{2-3}$  In particular,
gauge theories for the unconventional
metallic states found in the $t-J$ model,
 which is considered to be the simplest
model containing the essential strong
 correlation physics of the oxide
superconductors,$^4$ successfully account
for many of the unconventional
transport and collective mode
properties that are common to the
corresponding metallic phases
of these materials.$^{5,6}$
For example, the $T$-linear in-plane resistivity, as
well as the paradoxical observations of a hole-type Hall effect
in conjunction with a large Luttinger
Fermi surface, can be accounted for by such theories.

In a similar spirit,
analogous Ginzburg-Landau theories
of spin-charge separated superconductivity
itself have recently been proposed in
the literature.$^{7,8}$
It is presumed in such theories that
separate superfluid instabilities exists
 in both the spinon and holon systems
in isolation.  Here, the spinon superfluidity
is assumed to arise from singlet Cooper
pairing {\it \`a  la} the mean-field
resonating valence bond (RVB) scenario,$^9$
while the holon superfluidity is driven by
 Bose-Einstein condensation.  Given
the Ioffe-Larkin composition law, $R=R_b+R_f$,
which states that the physical
resistance is given by the sum
of the holon ($b$)
and spinon ($f$) contributions,
then true superconductivity occurs
 only when both species are superfluid.$^3$
However,
because such Ginzburg-Landau theories are
formulated in the continuum,
the analysis of the superconducting
transition in two dimensions is complicated
by the structure of the vortex cores.
 In addition, these
theories have been only analyzed in the
mean-field approximation, to date,
where fluctuations of the statistical
gauge-field related to the constraint
 against double occuppancy are neglected.
(Note that such excitations physically
represent so-called chiral spin fluctuations.$^{10}$)  As
a result, the precise shape that they obtain
for the phase boundary between the normal
 and the superconducting state is questionable.$^8$

In order to address the role played by
fluctuations of the statistical gauge-field,
we introduce here an
alternate spin-charge separated model of
strongly correlated superconductivity in
which only the phase of each order parameter is allowed to vary.
This model is a two-component
generalization of a lattice gauge theory
 model in two euclidean (1+1) dimensions
known as the Abelian-Higgs model.$^{11-13}$
Specifically, the theory contains a holon-pair
 phase order parameter (Higgs field)
and a spinon-pair phase order parameter
(Higgs field)
that couple in a gauge-invariant way to
the fluctuating
$U(1)$ (Abelian) statistical gauge-field
describing chiral spin fluctuations.$^{10}$
We henceforth
refer to the latter as the two-component
Abelian-Higgs (AH$^2$) model.$^{14}$
By considering both the continuum and the
 low temperature limits of this model,
we find that it exhibits a normal-to-superconducting
transition of the
Berezinskii-Kosterlitz-Thouless (BKT)
type in strictly two spacial dimensions.$^{15,16}$
Note that the low-temperature analysis is achieved
by a Villain duality transformation
of this model,$^{17}$ which has been
successfully employed in the past to study
 the one-component Abelian Higgs model and
the XY model.$^{11,18,19}$  We obtain the
following two major results: that ({\it i})
 the phase correlations corresponding
to the statistical gauge-field are short-range
at all temperatures, and that
({\it ii}) only phase correlations
corresponding to the physical
electronic order parameter
 show algebraic long-range order
below a BKT-type transition temperature,
$T_c\cong {\pi\over 2} (J_b^{-1}+J_f^{-1})^{-1}$,
 where $J_b$ and $J_f$
are the respective local phase-stiffnesses
of the holon-pairs and the spinon-pairs.
The former result ({\it i}) implies
 that statistical gauge-field
fluctuations do not acquire a gap
via the Higgs mechanism.  The latter
result ({\it ii}), on the other hand, implies
that this spin-charge separated model shows a
true superconducting transition
at $T_c$.$^{16}$  It is important to remark that
 the present calculation, which incorporates
fluctuations of the statistical gauge-field,
results in a transition temperature inferior
to the mean-field approximation result,$^{3,8}$
$T_c^{(0)}={\pi\over 2}\, {\rm min}(J_b,J_f)$,
as expected.
In addition, the presently obtained transition
 temperature yields a metal-superconductor
phase diagram as a function of hole
doping that qualitatively resembles
 that of the oxide superconductors
(see Fig. 1).$^{20}$ It is also argued that this
transition falls into the same universality class
as that of the XY model.  We thus predict that the
present  AH$^2$ model for spin-charge separated
 superconductivity has a universal jump in the
superfluid density at the transition.$^{16}$
Last, we compute the Wilson loop and find that
it shows a perimeter law in the superconducting phase.
On the other hand,
by continuity with the corresponding
results obtained for
the case of the one-component model,$^{11}$  we
conclude that the Wilson loop generally shows a ``confining''
area law in the normal phase.
This change in behavior simply
reflects the vortex binding-unbinding
transition.
We then argue, however, that the latter
``confinement'' effect is a trivial result
of ``electromagnetism'' in one space and
one time dimension (1+1), and hence that
fluctuations  in the statistical
gauge-field remain gapless in the normal phase.$^{21}$
This, hence, provides a basis
for the calculations of linear-in-$T$
resistance in the ``strange'' metallic
phase of the $t-J$ model,$^{5,6}$
which rely on this property.  Note that
gapless statistical gauge-field excitations
have also been shown to exist
in a two-component anyon superconductor
saddle-point of the $t-J$ model
known as the commensurate flux-phase
at low temperature and near half-filling.$^{22,23}$
Yet once dynamical effects are included, an
exponentially small gap appears close to
 half-filling in such case.$^{22}$

The remainder of the paper is organized as
follows.  In the next section we introduce
 the AH$^2$ model in the context of
strongly interacting electrons in
two dimensions.  Section III contains
a discussion of various limits of the
model, including the continuum limit.
 In section IV we study the phase
correlations in the low-temperatue
limit via the Villain duality transformation, while the nature of the
statistical gauge-field excitations
are treated in section V within the same context.
A renormalization group analysis of
the BKT-transition found in this
model is presented in section VI,
while we compare the theoretically
obtained phase diagram with that
observed experimentally in the oxide
superconductors as function of hole doping in section VII.
The nature of the normal state of the present model
is also discussed here.  Last, the phase correlators
of the model are computed in the ``spin-wave''
approximations in Appendix A, and those of the
one-component model are computed in the
low-temperature limit in Appendix B.

\bigskip
\bigskip
\centerline{\bf II. Two-component Abelian-Higgs Model}
\bigskip

Before introducing the Abelian-Higgs model
 for spin-charge separated superconductivity
in two dimensions, let us first consider the
corresponding normal state, which we take to
 be the unconventional metallic state found
in strongly interacting two-dimensional (2D)
electron systems first discussed by Ioffe
and Larkin in terms of the spin-charge
separated language of spinons and holons.$^{2,3}$
The archetypical theoretical
description of strong correlation physics
 in the context of the oxide superconductors
is given by  the  $t-J$ model defined over
the square lattice,$^4$ notably where double
 occuppancy of electrons at each site
is excluded as a result such
correlations.   This constraint may be
 imposed by the introduction
of an auxilliary slave-boson field, $b_i$,
 such that the electron field
is re-expressed as
$$c_{i\sigma}\rightarrow c_{i\sigma}b_i^{\dag}, \eqno (1)$$
along with the constraint
$c_{i\sigma}^{\dag}c_{i\sigma}+b_i^{\dag}b_i=1$
at each site.$^{22}$  Certain mean-field treatments
of the latter constraint
allowing the slave-boson to propagate result in a
spin-charge separated
description of the $t-J$ model, with $c_{i\sigma}$
representing the spinon field
and $b_i$ representing the holon field.$^{3-6}$
A statistical gauge field, $A_{\mu}$,
generated by the internal gauge symmetry
$(c_{i\sigma}, b_i)\rightarrow
(e^{i\theta_i}c_{i\sigma}, e^{i\theta_i}b_i)$ shown in Eq. (1)
then mediates interactions between the two species.
Within this representation, the ``strange''
metal is then characterized by a normal spinon
Fermi-liquid  state with a large Luttinger Fermi
 surface in conjunction with a normal (uncondensed)
 holon liquid state.
The statistical gauge-field  mediates
interactions among both species.$^{5,6}$
In particular, interactions of these
fluctuation with the uncondensed holon
liquid ultimately give rise to the $T$-linear prediction
for the resistance characteristic of the
``strange'' metal phase.$^5$

Upon lowering temperature, however,
meanfield resonanting-valence-bond (RVB)
studies of the $t-J$ model show that amplitude
 develops for an order parameter describing
a paired spinon state.$^{9}$  In addition,
amplitude for a separate  order parameter
representing the Bose-Einstein holon
condensate exists at low-temperature.
Hence by definition, in the mean-field
approximation, the unconventional metallic
state can only exist at temperatures above
both the highest temperature, $T_f$, at
which any spinon Cooper pairing instability
sets in, and above the Bose-Einstein condensation
temperature, $T_b$, for the holons (see Fig. 1).
Therefore, as temperature, $T$, is lowered, the
following three situations may arise:
({\it i}) $T_f<T<T_b$; ({\it ii}) $T_b<T<T_f$;
and ({\it iii}) $T<T_b,T_f$.   Given the Ioffe-Larkin
composition law,$^3$
$R=R_b+R_f$,
which states that the physical resistance is given by
the sum of the holon ($b$)
and spinon ($f$) contributions, then only regime ({\it iii})
is superconducting.
This implies that the critical temperature for the superfluid
transition is given by$^3$
$$T_c^{(0)}={\rm min}(T_b,T_f)\eqno (2)$$
 at the mean-field level.
 As has been discussed in the literature,$^{8}$ regimes ({\it i})
and ({\it ii}) are non-superconducting phases
that correspond, respectively, to a normal Fermi-liquid state
and a to spin-gap state.  The various regimes mentioned above
 are shown in Fig. 1.

The latter phase diagram  for strongly interacting electrons
in two dimensions is obtained in the mean-field approximation,
however, where fluctuations of the statistical gauge-field are
absent.   In this paper, we shall study what  effect such
fluctuations have on the previously discussed superfluid
transitions occurring in the spin-charge separated metallic
state by focusing on the phase degrees of freedom alone of
both the spinon and the holon superfluid order parameters.
In particular, consider
the following partition function for the two-component square
lattice Abelian-Higgs model,$^{14}$ which we presume models
the superfluid sector of the spinon/holon system in the presence
of statistical gauge-field
fluctuations:
$$Z=\int{\cal D}\phi_b(r){\cal D}\phi_f(r^{\prime})
{\cal D}A_{\mu}(r^{\prime\prime})
{\rm exp} (-E/T),\eqno (3)$$
where
$$\eqalignno{{E\over T} = & \beta_b\sum_{r,\mu}
\{1-{\rm cos}[\Delta_{\mu}\phi_b(r)-qA_{\mu}(r)]\}+
\beta_f\sum_{r,\mu}
\{1-{\rm cos}[\Delta_{\mu}\phi_f(r)-qA_{\mu}(r)]\}+\cr
&+{1\over{2g^2}}\sum_{r,\mu,\nu}
\{1-{\rm cos}[\Delta_{\mu}A_{\nu}(r)
-\Delta_{\nu}A_{\mu}(r)]\}. & (4)\cr}$$
Here, $\phi_b(r)$ and $\phi_f(r)$
represent the respective phases of the
holon and spinon order parameters,
$A_{\mu}(r)=A_{\vec r,\vec r+\hat\mu}$
denotes the statistical
gauge-field,$^{3-8}$ and $\Delta_{\mu}$  denotes
the lattice difference operator$^{11}$ ($\mu=x,y$).
The first two terms above are the lattice
versions of the stiffness energies for the phase fluctuations,
 while the last term is the corresponding stiffness energy for
 the statistical gauge-field fluctuations.
Also,  $\beta_b=J_b/T$ and $\beta_f=J_f/T$,
where $J_b$ and $J_f$ denote the local phase stiffness of
the holon order parameter and the spinon order parameter,
respectively, while
$g^{-2}=\chi_d/T$, where $\chi_d$ represents the local
stiffness for
statistical gauge-field fluctuations.
(Throughout,
we set $k_B$, $\hbar$
and the lattice constant, $a$, to unity, and we sum over
repeated indices.)  Spin-charge separated treatments of the
$t-J$ model in two dimensions
give a diamagnetic susceptibility for fluctuations in the statistical
gauge-field
of $\chi_d=\chi_f+\chi_b$, where $\chi_f$ and $\chi_b$
are the
diamagnetic susceptibilities of each species.$^{5,6}$
Within the metallic
(fluxless) saddle-point of these treatments,
$\chi_f\sim J(1-x)$ and
$\chi_b\sim tT_b/T$, where $x$ denotes the
hole concentration and
$T_b\lsim tx$ denotes the ideal Bose-Einstein condensation
temperature.   Also, $J_b\sim t^{\prime} x$,
where $t^{\prime}\lsim t$ denotes the relevant
matrix element for hopping in the holon
liquid, and $J_f\sim J(1-x/x_0)$, where $x_0\lsim 1$
 denotes the critical hole concentration above which
 the spinon pairing instability is absent.$^{9}$
Finally, $q$ represents the charge per site of each species.
 (Throughout, we follow the notations used in refs.
11, 18, and 19.)  We remind the reader that Landau-Ginzburg
 versions of this model, formulated in the 2D continuum,
have already been discussed in the literature.$^{7,8}$

Before we begin the analysis of the AH$^2$ model,
a few remarks are in order.  First, since physical electrons
have no statistical charge by Eq. (1),
we have restricted our analysis (4) to
statistically charge neutral AH$^2$ models;
i.e.,  the statistical charge per site, $q$,
 of the holons and of the spinons
is {\it equal}.  Second,  we implicitly presume that
the spinon order parameter, $\phi_f$, results from Cooper pairing
of the spinons
{\it \`a la} the
RVB picture for superconductivity.$^{9}$
Hence, from this point on we shall take a charge
per site of $q=2$ for the present model, which implies that the
holon condensate is made up pairs of holons as well.  Note
that a holon pairing instability
consistent with the previous assignment
has been shown to exist in the
``strange'' metallic phase of the $t-J$ model,$^6$
which is the normal state of the present model.
Also, in
the limit near half-filling
relevant to the oxide superconductors,
the density of holon pairs is small
and the overlap between two pairs of bosons should be negligible.
Therefore, a bose condensate of dilute bosonic molecules
can form,$^{24}$
resulting in the holon order parameter, $\phi_b$, above.
If, on the other
hand, the holon condensate where to result from a
condensation of unpaired
holons as is suggested by certain mean-field
treatments,$^{4,5,9}$ then such an order parameter
must necessarily be defined on a lattice
different from that of the spinon order parameter
by the requirement of statistical charge
neutrality (1).  The latter physical situation,
however, cannot be described by the present AH$^2$ model,
which has only one underlying lattice.

\bigskip
\bigskip
\centerline{\bf III. Limits}
\bigskip
The nature of the above AH$^2$ model (4) for spin-charge separated
superconductivity on the square lattice can be uncovered,
in large part, by an analysis
of various limits, as we discuss below.

{\it A. Continuum Limit.}  Consider the limit
where $q^2g^2\beta_b,\, q^2g^2\beta_f\ll 1$ and where
fluctuations in the fields are at longwavelength,
which corresponds to the continuum limit [see Eqs. (4) and (20)].
 In this case, the last term in the energy functional (4)
is negligible, whereas the first two terms may approximated
by
$${E\over T} =  {1\over 2}\int d^2r[\beta_b(\vec\nabla\phi_b-q\vec
A)^2+\beta_f(\vec\nabla\phi_f-q\vec A)^2]. \eqno (5)$$
As is customary in the treatment of the XY model, let us now
 separate the phase configurations into
``spin-wave'' and vortex components;$^{19}$ i.e., let
$\phi_s=\phi_s^{(w)}+\phi_s^{(v)}$ for each species $s=b,f$, where
$\phi_s^{(w)}$ represents the smooth portion of the
configuration such that
$\vec\nabla\times \bigl(\vec\nabla\phi_s^{(w)}\bigr)=0$,
 and where $\phi_s^{(v)}$
represents the portion of the configuration containing vortex
 singularities such that
$\vec\nabla\cdot\bigl(\vec\nabla\phi_s^{(v)}\bigr)=0$.
Taking the Coulomb gauge, $\vec\nabla\cdot\vec A=0$, we see
that statistical gauge-field fluctuations only communicate
with the component of the phase configuration containing
vorticity, since the statistical  gauge-field is purely
tranverse in this gauge, as are such configurations.
Therefore, integrating out first the gauge-field
excitations in the corresponding partition function
(3) within this gauge leaves us with following
effective action for the vortex component:
$${E_v\over T} =  {\bar\beta\over 2}
\int d^2r\bigl(\vec\nabla\phi_ f^{(v)}-\vec\nabla\phi_b^{(v)}\bigr)^2,
 \eqno (6)$$
where
$$\bar\beta=(\beta_b^{-1}+\beta_f^{-1})^{-1}.\eqno (7)$$
The remaining smooth ``spin-wave'' components result
in a trivial Gaussian action,
$${E_w\over T} =  {1\over 2}
\int d^2r\bigl[\beta_b\bigl(\vec\nabla\phi_
b^{(w)}\bigr)^2+\beta_f\bigl(\vec\nabla\phi_ f^{(w)}\bigr)^2\bigr].$$
Hence, by eqs. (6) and (7) we expect a BKT
transition to occur in the physical
electronic phase,
$$\phi_{\rm el}=\phi_f-\phi_b\eqno (8)$$
when $2\pi\bar\beta=4$, which implies a
corresponding BKT transition temperature of
$$T_c\cong{\pi\over 2}(J_b^{-1}+J_f^{-1})^{-1}.\eqno (9)$$
In the continuum limit, therefore,  statistical gauge-field
 fluctuations suppress, but do not destroy, the
 superconductivity in this spin-charge separated
 system in relation to the Ioffe-Larkin mean-field
approximation result; i.e. $0<T_c<{\rm min} (T_b,T_f)$,
where $T_b={\pi\over 2} J_b$ and $T_f={\pi\over 2} J_f$
 are the corresponding transition temperatures of each
specie separately.  As we will show in the next section, this result is
generally valid outside of the continuum limit, as well.

{\it B. XY model + Pure Gauge Theory.}  Consider now
the limit where $\beta_b$ (or $\beta_f$) $\gg 1$.  After making the
gauge-transformation
$A_{\mu}^{\prime}=A_{\mu}-q^{-1}\Delta_{\mu}\phi_b$,
we can rewrite the energy functional (4) as
$$\eqalignno{{E\over T} = & \beta_b\sum_{r,\mu}
[1-{\rm cos}(qA_{\mu}^{\prime})]+
\beta_f\sum_{r,\mu}
\{1-{\rm cos}[\Delta_{\mu}\phi_{\rm el}(r)-qA_{\mu}^{\prime}(r)]\}+\cr
&+{1\over{2g^2}}\sum_{r,\mu,\nu}
\{1-{\rm cos}[\Delta_{\mu}A_{\nu}^{\prime}(r)
-\Delta_{\nu}A_{\mu}^{\prime}(r)]\}. & (10)\cr}$$
Hence, the present limit imposes the constraint
$A_{\mu}^{\prime}=2\pi n/q$, where $n$ is any integer.  The above energy
functional then reduces to
$${E\over T} \rightarrow
\beta_f\sum_{r,\mu}
\{1-{\rm cos}[\Delta_{\mu}\phi_{\rm el}(r)]\}+
{1\over{2g^2}}\sum_{r,\mu,\nu}
\{1-{\rm cos}[\Delta_{\mu}A_{\nu}^{\prime}(r)
-\Delta_{\nu}A_{\mu}^{\prime}(r)]\}. \eqno (11)$$
The latter expression, in conjunction with the former
 constraint, contains an XY model that is decoupled from a
 pure gauge theory known in the literature as
$Z_q$.$^{13}$  The XY model part implies a BKT transition
at $T_f$ for the physical electronic phase, $\phi_{\rm el}$,
which is consistent with
the results (9) obtained previously in the continuum limit
for the present case, $J_b\rightarrow\infty$. The $q$-state ($Z_q$)
pure gauge theory part of the energy functional (11),
 on the other hand,
can be easily shown to exhibit no phase transition
in two dimensions by choosing to work in the Landau gauge
 ($A_x=0$).$^{13}$
This, of course, reduces the theory to one dimension.

{\it C. One-component Abelian-Higgs Model.}  Last, let us
 take the limit
$\beta_b=0$ or $\beta_f=0$, which corresponds to the well
studied one-component Abelian-Higgs model. The latter model
sustains no Higgs phenomenon, since the long-range interaction is
 screened, and in general, it is thought to be free of phase
transitions in two dimensions.$^{11-13}$
In fact, the phase correlation length can be shown to be
 finite in the low-temperature limit (see Appendix B).
Again, these results are consistent with those obtained
in the continuum limit (9); i.e., $T_c=0$ in the case
that $J_b=0$ or $J_f=0$.

In summary, we find that the physical electronic phase (8)
undergoes a BKT transition at a critical temperature given
by expression (9), while the
gauge-field excitations experience no phase transitions at
all, in the limits considered above.   Below, we show that
these findings are corroborated by a low-temperature analysis
 of the model.

\bigskip
\bigskip
\centerline{\bf IV. Phase Correlations}
\bigskip

With the intent of understanding the superconducting
properties of our model (4) for spin-charge separated
superconductivity in two dimensions, let us consider
now the low-temperature limit $\beta_b,\beta_f, g^{-2}\gg 1$.
In this limit, the most important configurations of
both the phases and the statistical gauge-field lie
at extrema of the cosine functions found in expression (4).  Thus,
we may employ well-known
Villain duality transformations
that are successful in analysing
 both the XY
model$^{18,19}$ and the one-component Abelian-Higgs model$^{11}$
 in the same limit.  This approximation amounts to
replacing the exponential of a cosine function by the
 sum of exponentials of parabolas, each situated at
the appropriate extrema of the cosine function.$^{17}$
Below, we show that the latter analysis leads to a
generalized Coulomb gas
ensemble that exhibits a BKT phase transition
consistent with the discussion found in the previous section
(III.A).  Note also that the analogous treatment of
 the phase auto-correlations function in the special
 case of the one-component Abelian-Higgs model
 ($\beta_b=0$ or $\beta_f=0$), which is useful
in the study of the unconventional normal state
 corresponding to the present model for spin-charge separated
superconductivity,$^{21}$ is found in Appendix B.

{\it A. Coulomb Gas Representation.} Application of
 the above mentioned low-temperature approximation
to the presently considered AH$^2$ model reduces to
 the substitution of the
mathematical identity
$$e^{-\beta(1-{\rm cos}\,\theta)}\cong (2\pi\beta)^{-1/2}
\sum_{n=-\infty}^\infty
e^{in\theta}e^{-n^2/2\beta},\eqno (12)$$
that is valid in the limit $\beta\rightarrow\infty$,
into the partition function (3) corresponding to the
energy functional (4).  By closely following the analogous
treat of the one-component version of our model
 discussed in ref. 11, and  performing the integrals
that remain over the fields $\phi_b$, $\phi_f$, and $A_{\mu}$,
 we ultimately arrive at the following ``roughening model''
representation for this partition function:
$$\eqalignno{Z=\sum_{\{n_b(r)\}}\sum_{\{n_f(r)\}}
{\rm exp}\Biggl\{ & -{1\over{2\beta_b}}
\sum_{r,\mu}[\Delta_{\mu}n_b(r)]^2
-{1\over{2\beta_f}}\sum_{r,\mu}
[\Delta_{\mu}n_f(r)]^2\cr
&-{(qg)^2\over 2}\sum_r[n_b(r)+n_f(r)]^2\Biggr\},
& (13)\cr}$$
where $n_b(r)$ and $n_f(r)$ are integer fields
that lie on the dual lattice $r$.$^{19}$
Also, after application of the Poisson summation
formula, this expression may be
transformed further into
$$\eqalignno{Z=&\int_{-\infty}^{\infty}\Pi_rd\theta_b(r)
\int_{-\infty}^{\infty}\Pi_rd\theta_f(r)
\sum_{\{q_b(r)\}}\sum_{\{q_f(r)\}}
{\rm exp}\Biggl\{-{1\over{2\beta_b}}\sum_{r,\mu}(\Delta_{\mu}\theta_b)^2
\cr
&-{1\over{2\beta_f}}\sum_{r,\mu}(\Delta_{\mu}\theta_f)^2
-{(qg)^2\over 2}\sum_r(\theta_b+\theta_f)^2+2\pi
i\sum_r(q_b\theta_b+q_f\theta_f)\Biggr\}, & (14)\cr}$$
where $q_b(r)$ and $q_f(r)$ are the respective
dual lattice vortex ``charges''
of each specie that range over the integers.
After integrating out the fields, $\theta_b(r)$
and $\theta_f(r)$, that result from the Poisson
 summation formula, we find that the present partition
function is equal to the product
of a trivial gaussian factor with the following
generalized 2D
Coulomb gas ensemble average:
$$Z_{\rm Coulomb}=\sum_{\{q_{\alpha}(r)\}}{\rm
exp}\Biggl[-(2\pi)^2\bar\beta{1\over
2}\sum_{r,r^{\prime}}q_{\alpha}(r)G_{\alpha\beta}
(r,r^{\prime})q_{\beta}(r^{\prime})\Biggr],\eqno (15)$$
where
$$G_{\alpha\beta}(r,r^{\prime})=(2\pi)^{-2}
\int_{\rm BZ}d^2k
\, e^{i\vec k\cdot(\vec r-\vec r^{\prime})}
G_{\alpha\beta}(\vec k)\eqno (16)$$
 is the Greens function, such that the matrix
inverse of its Fourier transform, $G(\vec k)$, is
$$G_{\alpha\beta}^{-1}(\vec k)=
{\bar\beta\over{\beta_{\alpha}}}
\tilde k^2\delta_{\alpha\beta}+\bar\beta q^2g^2
u_{\alpha}u_{\beta}.\eqno (17)$$
The above indices $\alpha$ and $\beta$
represent the internal spin/charge label,
$b$ or $f$, while $\tilde k^2=4-2{\rm cos}
\, k_x -2{\rm cos}\, k_y$ are the
 eigenvalues of the lattice Laplacian operator.
In addition, $\vec u=(1,1)$ represents
 the spin/charge component of the statistical degree
of freedom [see Eq. (25)].
  After inverting the righthand side of Eq. (17),
we find that the matrix Greens function is decomposable
into  long-range and short-range components
$$G_{\alpha\beta}(\vec k)=G_{\alpha\beta}^{({\rm lr})}(\vec
k)+G_{\alpha\beta}^{({\rm sr})}(\vec k), \eqno (18)$$
such that
$$\eqalignno{G_{\alpha\beta}^{({\rm lr})}(\vec
k)=&v_{\alpha}v_{\beta}{1\over{\tilde k^{2}}}, & (19a)\cr
G_{\alpha\beta}^{({\rm sr})}(\vec
k)=&{\beta_{\alpha}\beta_{\beta}\over{\beta_b\beta_f}}
{1\over{\tilde k^2+\lambda_{\rm st}^{-2}}}, & (19b)\cr}$$
where $\vec v=(1,-1)$ represents the spin/charge components
 of the physical electronic  degree of freedom (8), and where
$$\lambda_{\rm st}^{-2}=
(qg)^2(\beta_b+\beta_f)\eqno (20)$$
yields the characteristic length scale,
$\lambda_{\rm st}$, for fluctuations of
the statistical gauge-field.
The conjunction of these last few expressions
(Eqs. 15-20) are, in effect, the Coulomb gas
representation of the AH$^2$ model.

The above formulae may be reduced further, however.
 After substituting in the long-range and short range
 components of the matrix Greens function discussed
 above into (15), algebraic manipulation then yields
 that this generalized Coulomb gas partition function
 may be expressed as
$$Z_{\rm Coulomb}=\sum_{\{q_b(r)\}}
\sum_{\{q_f(r)\}}{\rm exp}\Biggl\{2\pi\bar\beta
\sum_{(r,r^{\prime})}
[\Gamma_{\rm lr}(\vec r-\vec r\,^{\prime})
q_{\rm el}(r)q_{\rm el}(r^{\prime})+
\Gamma_{\rm sr}(\vec r-\vec r\,^{\prime})
q_{\rm st}^{\prime}(r)q_{\rm st}^{\prime}(r^{\prime})]\Biggr\},
\eqno (21)$$
where the physical electronic
(el) flux-charge and the {\it modified}
statistical (st) flux-charge are given respectively by
$$\eqalignno{q_{\rm el}(r)=&q_f(r)-q_b(r),&(22)\cr
     q_{\rm st}^{\prime}(r)=
&\Biggl({\beta_f\over{\beta_b}}\Biggr)^{1/2}q_f(r)
+\Biggl({\beta_b\over{\beta_f}}\Biggr)^{1/2}q_b(r),&(23)\cr}$$
and where $(r,r^{\prime})$
denote combinations
of points covering the dual lattice.$^{25}$
Corresponding to these charges are long-range (lr)
and short-range (sr) potentials given respectively by
$$\eqalignno{\Gamma_{\rm lr}(\vec r)
=&\int_{\rm BZ}{d^2k\over{2\pi}}
(1-e^{i\vec k\cdot\vec r}){1\over{\tilde k^2}},&(24a)\cr
             \Gamma_{\rm sr}
(\vec r)=&\int_{\rm BZ}{d^2k\over{2\pi}}
(1-e^{i\vec k\cdot\vec r})
{1\over{\tilde k^2+\lambda_{\rm st}^{-2}}}.&(24b)\cr}$$
Note that for the sake of technical conveniance,
we have added an overall constant to the matrix
Greens function represented by the first term
in the above two equations.  Due to the existence
 of a long range force, charge nuetrality
is enforced for each species  and this constant
term therefore sums  to zero for all of the
relevant charge configurations in the energy
 functional of the enseble average [see Eq. (15)].
Since the effective Coulomb gas inverse
temperature scale is given by $\bar\beta$,
we expect there to exist a
 binding-unbinding transition of the
BKT type modified by short-range
interactions for the physical electronic
flux-charges, $q_{\rm el}(r)$, at a corresponding
critical temperature.
On the other hand, for the case of purely
statistical flux-charge
configurations, $q_{\rm el}(r)=0$
 and $q_{\rm st}(r)=q_f(r)+q_b(r)\neq 0$,
Eq. (21) indicates that there exists
only short-range forces, where $\lambda_{\rm st}$
determines the length scale for such
forces in units
of the lattice spacing, $a$.
  Thus, we expect
{\it no} BKT transition in such a case.
Below, we elaborate on the
consequences of the latter statements
to show that ({\it i}) a 2D
superconducting transition of the
BKT type exists {\it only}
for the physical
electronic phase (8),
and that ({\it ii}) all other phase auto-correlations,
including those among the
statistical phase,
$$\phi_{\rm st}=\phi_f+\phi_b,\eqno (25)$$
are short-range at {\it all} temperatures.
Note that exactly the opposite behavior is
exhibited by two square-lattice XY
models stacked up on top of each other with
nearest-neighbor coupling, where the phase difference
between layers has short-range order at zero temperature, while
the sum of the phases between layers shows quasi long-range order
at low temperature.$^{26}$

{\it B. Auto-correlation Functions.}  We now wish
to probe the phase correlations of the presently
considered AH$^2$ model for spin-charge separated
 superconductitvity in two dimensions.  The
gauge-invariant phase correlation function
may be expressed as
$$C_{12}=
\Biggl\langle {\rm exp}\Biggl\{
i\sum_{\alpha}p_{\alpha}\Biggl[\phi_{\alpha}(\vec r_1)-
\phi_{\alpha}(\vec r_2)+q\int_{\vec r_1}^{\vec
r_2}A_{\mu}dr_{\mu}\Biggr]\Biggr\}
\Biggr\rangle
 = {Z^{\prime}\over Z},\eqno (26)$$
($\alpha=f,b$)
where the partition function $Z^{\prime}$ differs from $Z$
only by the addition of the exponent
$-i\sum_{\alpha}p_{\alpha}[\phi_{\alpha}(\vec r_1)-
\phi_{\alpha}(\vec r_2)+q\int_{\vec r_1}^{\vec r_2}A_{\mu}dr_{\mu}]$
to the energy functional (4).  Again, following
refs. 18 and 19, $Z^{\prime}$
may be computed via the Villain
duality transformation.  In particular,
substitution of the mathematical identity
(12) that is valid in the low-temprature
 limit ultimately yields the following ``roughening model''
representation for $Z^{\prime}$:
$$\eqalignno{Z^{\prime}=\sum_{\{n_b(r)\}}\sum_{\{n_f(r)\}}
{\rm exp}\Biggl\{ & -{1\over{2\beta_b}}
\sum_{r,\mu}[\Delta_{\mu}n_b(r)+p_b\eta_{\mu}(r)]^2
-{1\over{2\beta_f}}
\sum_{r,\mu}[\Delta_{\mu}n_f(r)+p_f\eta_{\mu}(r)]^2\cr
&-{(qg)^2\over 2}\sum_r[n_b(r)+n_f(r)]^2\Biggr\},& (27)\cr}$$
where the dual lattice ``dipole'' vector,
$\eta_{\mu}(r)$, has a value of
$\pm 1$ if it intersects a fixed path
on the original lattice connecting
points $\vec r_1$ and $\vec r_2$,
and it vanishes otherwize.
As before, after the application of the
Poisson summation formula, and subsequently
integrating over the fields that this formula generates, we find
that
$$\eqalignno{Z^{\prime}=&{\rm exp}\Biggl[-{1\over
2}\Biggl({p_b^2\over{\beta_b}}+
{p_f^2\over{\beta_f}}\Biggr)\sum_{r,\mu}
\eta_{\mu}^2(r)\Biggr]\times\cr
&\times\sum_{\{q_{\alpha}(r)\}}{\rm exp}
\Biggl\{-(2\pi)^2\bar\beta{1\over 2}\sum_{r,r^{\prime}}[q_{\alpha}(r)+
q_{\alpha}^{\prime}(r)]
G_{\alpha\beta}(r,r^{\prime})
[q_{\beta}(r^{\prime})+q_{\beta}^{\prime}
(r^{\prime})]\Biggr\},& (28)\cr}$$
where the external vortex ``charges''
 corresponding to each species are given by
$$q_{\alpha}^{\prime}(r)=-{p_{\alpha}\over{2\pi i\beta_{\alpha}}}
\eta(r),\eqno (29)$$
with $\eta(r)=\Delta_{\mu}\eta_{\mu}(r)$.$^{18,19}$
This expression constitutes the Coulomb gas representation
 of the modified partition function $Z^{\prime}$.

The above expression for the modified partition function
can be reduced further by substituting in the explicit
form of the matrix Greens function $G$ [see Eqs. (16-19)].
We then find that the preceeding correlation function,
 $Z^{\prime}/Z$, may
be factorized into
$$C_{12}
\cong G_{\rm wave}(\vec r_1-\vec r_2)
G_{\rm Coulomb}(\vec r_1-\vec r_2), \eqno (30)$$
where
$$\eqalignno{G_{\rm wave}(\vec r_1-\vec r_2)
={\rm exp}\Biggl\{&-{1\over 2}\Biggl({p_b^2\over{\beta_b}}
+{p_f^2\over{\beta_f}}\Biggr)\sum_{r,\mu}\eta_{\mu}^2(r)\cr
&-{1\over{\pi}}{\bar\beta\over
{\beta_b\beta_f}}\sum_{r,r^{\prime}}
\eta(r)[p_{\rm el}^{\prime 2}
\Gamma_{\rm lr}(\vec r-\vec r\,^{\prime})
+p_{\rm st}^2\Gamma_{\rm sr}
(\vec r-\vec r\,^{\prime})]\eta(r^{\prime})
\Biggr\}& (31)\cr}$$
and
$$\eqalignno{G_{\rm Coulomb}
(\vec r_1-\vec r_2)=
\Biggl\langle{\rm exp}\Biggl\{&i{2\bar\beta\over
{(\beta_b\beta_f)^{1/2}}}\sum_{r,r^{\prime}}
[p_{\rm el}^{\prime}q_{\rm el}(r)
\Gamma_{\rm lr}(\vec r-\vec r\,^{\prime})
+\cr
&+p_{\rm st}q_{\rm st}^{\prime}(r)
\Gamma_{\rm sr}(\vec r-\vec r\,^{\prime})]
\eta(r^{\prime})\Biggr\}\Biggr\rangle_{\rm Coulomb},&(32)\cr}$$
with $p_{\rm el}^{\prime}={1\over
2}[(\beta_b/\beta_f)^{1/2}p_f-(\beta_f/\beta_b)^{1/2}p_b]$
and $p_{\rm st}={1\over 2}(p_f+p_b)$.
After some manipulation (see Appendix A), it can be shown that
in the limit $|\vec r|\gg\lambda_{\rm st}$,
$$G_{\rm wave}(\vec r)=
{\rm exp}\Biggl[-{p_{\rm st}^{\prime\prime 2}\over{2\bar\beta}}
(r+\pi^{-1}l{\rm n}\lambda_{\rm st})\Biggr]
{\rm exp} \Biggl[-{p_{\rm el}^{\prime\prime 2}\over{2\bar\beta}}
{\Gamma_{\rm lr}(r)
\over{\pi}}\Biggr],\eqno (33)$$
with
$$\eqalignno{
p_{\rm el}^{\prime\prime}&={2\bar\beta\over{(\beta_b\beta_f)^{1/2}}}
p_{\rm el}^{\prime}=p_{\rm
el}+\bar\beta\Biggl({1\over{\beta_f}}-{1\over{\beta_b}}\Biggr)
 p_{\rm st},& (34a)\cr
p_{\rm st}^{\prime\prime}&={2\bar\beta\over{(\beta_b\beta_f)^{1/2}}}
p_{\rm st}, & (34b)\cr}$$
and where $p_{\rm el}={1\over 2}
(p_f-p_b)$ and $l{\rm n}\,\lambda_{\rm st}=(2\pi)^{-1}
\int_{\rm BZ}d^2k
(\tilde k^2+\lambda_{\rm st}^{-2})^{-1}$.
Hence, {\it all} phase correlations
are short-range at all temperatures
unless $p_{\rm st}=0$.
 In particular,  purely statistical phase
correlations, with $p_{\rm el}=0$ and $p_{\rm st}\neq 0$,
are short-range.  This implies that, within the present model,
transverse statistical gauge-field
fluctuations do not acquire a gap
via a Higgs mechanism.

A short-range order to algebraic long-range
order transition of the BKT type does, however,
exist for purely electronic
phase correlations, $p_{\rm st}=0$ and
$p_{\rm el}\neq 0$, as we demonstrate below.
But before considering this case in particular, let us first
compute
$G_{\rm Coulomb}$
for the presently considered AH$^2$ model on the
 square-lattice in general.  This function may be
calculated by following
well-known methods that have been successfully
 employed in the corresponding computation
for the case of the XY model.$^{18,19}$  In particular,
given the definitions
$$\eqalignno{\sigma_{\rm lr}(r)&=
\sum_{r^{\prime}}\Gamma_{\rm lr}
(\vec r-\vec r^{\prime})\eta(r^{\prime}), & (35a)\cr
\sigma_{\rm sr}(r)&=\sum_{r^{\prime}}\Gamma_{\rm sr}(\vec r-\vec
r^{\prime})\eta(r^{\prime}), & (35b)\cr}$$
we can re-express the latter correlation function (32) as
$$G_{\rm Coulomb}
(\vec r_1-\vec r_2)=
\Biggl\langle{\rm exp}\Biggl\{i{2\bar\beta\over
{(\beta_b\beta_f)^{1/2}}}\sum_{r}
\langle[p_{\rm el}^{\prime}q_{\rm el}(r)
\sigma_{\rm lr}(r)+p_{\rm st}q_{\rm st}^{\prime}(r)
\sigma_{\rm sr}(r)]\rangle\Biggr\}
\Biggr\rangle_{\rm Coulomb}. \eqno (36)$$
Since charge correlations in the present
 model are strong as a result of the
existence long-range forces, we can
make the following cummulant expansion:
$$\eqalignno{G_{\rm Coulomb}
(\vec r_1-\vec r_2)&=
{\rm exp}\Biggl\{-{1\over 2}{4\bar\beta^2\over
{\beta_b\beta_f}}\sum_{r^{\prime},r^{\prime\prime}}
\langle[p_{\rm el}^{\prime}q_{\rm el}(r^{\prime})
\sigma_{\rm lr}(r^{\prime})+p_{\rm st}q_{\rm st}^{\prime}(r^{\prime})
\sigma_{\rm sr}(r^{\prime})]\cr
&\ \ \ \ \ \ \ \ \ \ \ \ \ \ \ \ \ \ \ \ \ \ \ \ \times [p_{\rm
el}^{\prime}q_{\rm el}(r^{\prime\prime})
\sigma_{\rm lr}(r^{\prime\prime})+p_{\rm st}q_{\rm
st}^{\prime}(r^{\prime\prime})
\sigma_{\rm sr}(r^{\prime\prime})]\rangle\Biggr\}. & (37)\cr
}$$
But since only the lowest energy vortex-antivortex
excitations need be accounted for at the low
temperatures presently considered, the
above expression then reduces to
$$G_{\rm Coulomb}
(\vec r_1-\vec r_2)=
{\rm exp}\Biggl[-{1\over 2}\sum_{r^{\prime},r^{\prime\prime}}\langle
q_{i}(r^{\prime})q_{i}(r^{\prime\prime})\rangle
\sigma (r^{\prime})\sigma (r^{\prime\prime})\Biggr],
\eqno (38)$$
where
$$\sigma(r^{\prime})=
{\rm sgn}\,(\beta_b-\beta_f)\,p_{\rm el}^{\prime\prime}
\sigma_{\rm lr}(r^{\prime})+{\rm
min}\,[(\beta_b/\beta_f)^{1/2},(\beta_f/\beta_b)^{1/2}] \,p_{\rm
st}^{\prime\prime}
\sigma_{\rm sr}(r^{\prime}),\eqno (39)$$
and where
the Coulomb charge correlation function for the species labeled by $i$,
such that $\beta_i={\rm min}\,(\beta_b,\beta_f)$, is given by
$$\langle q_{i}(0)q_{i}(r)\rangle =
-{\rm exp}\Biggl\{-2\pi\bar\beta\Biggl
[\Gamma_{\rm lr}(r)+{\rm min}\Biggl({\beta_b\over{\beta_f}},
{\beta_f\over{\beta_b}}\Biggr)
\Gamma_{\rm sr}(r)\Biggr]\Biggr\}.\eqno (40)$$
Note, therefore, that vortex-antivortex excitations
 are the least energetically costly for the latter species.

In the particular case of the  physical electronic
phase correlator, for which
$p_{\rm st}=0$, the above expression (38) has
 precisely the same form as that corresponding
to the XY model;$^{18,19}$ i.e.,
$\sigma(r^{\prime})=\pm p_{\rm el}\sigma_{\rm lr}(r^{\prime})$.
 Due to the fact that vortices are bound to anti-vortices in the present
low-temperature limit, we may expand $\sigma_{\rm lr}(r^{\prime})$ and
$\sigma_{\rm lr}(r^{\prime\prime})$ with reference to the center of
mass coordinates; i.e.,
$$\eqalignno{\sigma_{\rm lr}(r^{\prime}) =
&\sigma_{\rm lr}(R)+{1\over 2}
\vec r\cdot\vec\nabla \sigma_{\rm lr}(R),\cr
    \sigma_{\rm lr}(r^{\prime\prime}) =
&\sigma_{\rm lr}(R)-{1\over 2}\vec r\cdot
\vec\nabla \sigma_{\rm lr}(R),\cr}$$
where $R$ is the center of mass coordinate and
 $r$ is the relative coordinate between points
 $r^{\prime}$ and $r^{\prime\prime}$.  Due to
 the charge neutrality condition effectively
imposed by the presence of long-range
interactions, only the last terms above
 remain in sum over coordinates found in the
exponent of expression (38).  This ultimately results in
$$G_{\rm Coulomb}
(\vec r_1-\vec r_2)=
{\rm exp}\Biggl[{1\over
2}p_{\rm el}^2\sum_{r}{1\over 2}r^2
\langle q_{i}(0)q_{i}(r)\rangle
\sum_{R}{1\over 4}[\vec\nabla\sigma_{\rm lr} (R)]^2\Biggr],
 \eqno (41)$$
where the factor of ${1/2}$ found in the sum over the
relative coordinate $r$ is due to an angle average.
 However, it can be shown in a straight forward manner that$^{18,19}$
$$\sum_{R}{1\over 4}[\vec\nabla\sigma_{\rm lr} (R)]^2
=\pi\Gamma_{\rm lr}(\vec r_1-\vec r_2).$$
Inserting this identity into (41), we arrive at the
final closed form expression $$G_{\rm Coulomb}
(\vec r_1-\vec r_2)=
{\rm exp}\Biggl[{1\over
4}\pi p_{\rm el}^2\Gamma_{\rm lr}(\vec r_1-\vec r_2)\sum_{r}r^2\langle
q_{i}(0)q_{i}(r)\rangle
\Biggr]. \eqno (42)$$
Last, factoring in above the previous result (33) for the ``spin-wave''
contribution, we find that
in the limit $|\vec r_1-\vec r_2|\gg\lambda_{\rm st}$,
the auto-correlation function (26) for the physical
electronic phase is given by
$$\Biggl\langle e^{ip_{\rm el}[\phi_{\rm el}(1)-
\phi_{\rm el}(2)]}\Biggr\rangle \cong {\rm exp}\Biggl[-{p_{\rm
el}^2\over{2\pi\beta_{\rm eff}}}
\Gamma_{\rm lr}(\vec r_1-\vec r_2)\Biggr]\eqno (43)$$
in the low-temperature limit,
with an effective temperature scale
$${1\over\beta_{\rm eff}}={1\over{\bar\beta}}
-{\pi^2\over 2}\sum_{|\vec r|>a}r^2\langle
 q_{i}(0)q_{i}(r)\rangle.
\eqno (44)$$
We now recall that $\Gamma_{\rm lr}(r)
\cong{\rm ln}(2^{3/2}e^{\gamma}r)$,
where $\gamma$ is Euler's constant.$^{19}$  Hence,
in the limit $\lambda_{\rm st}\rightarrow\infty$,
where statistical gauge-field
fluctuations in the energy functional (4)
are ``frozen out'',
inspection of Eq. (40) reveals that
 $\beta_{\rm eff}^{-1}$ diverges at
$2\pi\,{\rm min}(\beta_b,\beta_f)=4$.
This implies a BKT transition temperature
of $T_c\cong{\pi\over 2}\,{\rm min}(J_b,J_f)$
 below which algebraic
long-range order sets in (see dashed lines, Fig. 1).
The latter result is precisely what one expects
from employing the Ioffe-Larkin formula
 for the London penetration
length in a spin-charge separated
superconductor;$^{3,8}$ i.e., the mean-field approximation
 result (2).  On the other hand,
in the general case of {\it finite} $\lambda_{\rm st}$,
similar considerations reveal
that $\beta_{\rm eff}^{-1}$ diverges at
$2\pi\bar\beta=4$, which implies a
BKT transition temperature given by Eq. (9).
Hence, both the previous continuum limit analysis and the present
low-temperature analysis arrive at the same answer
concerning the nature of the BKT phase-transition in the AH$^2$ model.

Last, we note that the previous analysis for the Coulomb
gas factor, $G_{\rm Coulomb}$, in the particular case of
the physical electronic phase correlator ($p_{\rm st}=0$)
 is also valid for the general case in the limit of maximum
 statistical gauge-field fluctuations, $\lambda_{\rm st}\rightarrow 0$
[see Eq. (32)],$^{18,19}$ with the expception that
we must replace $p_{\rm el}$ in Eq. (42) with the
 orginal expression (34a) for $p_{\rm el}^{\prime\prime}$.
  Factoring in the latter result  with the previous result
(33) for the ``spin-wave'' contribution, one finds that
the correlation function (26) is given by
$$
C_{12}
\cong {\rm exp}\Biggl(-{p_{\rm st}^{\prime\prime 2}\over{2\bar\beta}}
|\vec r_1-\vec r_2|\Biggr)
{\rm exp}\Biggl[-{p_{\rm el}^{\prime\prime 2}\over{2\pi\beta_{\rm eff}}}
\Gamma_{\rm lr}(\vec r_1-\vec r_2)\Biggr].\eqno (45)$$
Hence, we see that in the limit of maximum fluctuations
in the statistical gauge-field considered here, while
the phase auto-correlations are generally short-range
, a remnant of the long-range
behavoir that exists in the special case of the
 electronic auto-correlations function  (43) persists.

\bigskip
\bigskip
\centerline{\bf V. Statistical Gauge-field
Excitations and the Wilson Loop}
\bigskip

We have demonstrated above that the statistical
gauge-field does not acquire a
gap via the Higgs mechanism, since
the corresponding auto-correlations for the
statistical phase (25) are short-range.
However, a gap in a $U(1)$ gauge-field can also result
from confinement effects in 2+1 dimension,$^{27}$ for example.
Such effects are conveniantly probed by the Wilson loop,
 $\langle{\rm exp}(ie^{\prime}\oint
A_{\mu}dx_{\mu})\rangle$.  Below we show
that the presently considered
AH$^2$ model displays a perimeter law
for this quantity at all temperatures
below the BKT-transition discussed above,
which simply reflects the existence of
bound vortices, and which corresponds to a
``deconfined'' phase.  Also, although we do
find that the Wilson loop exhibits a ``confining''
area law at temperatures above the BKT phase-transition,
we argue that this behavior is a general property
of {\it pure} Abelian gauge theories in two
euclidean dimensions.  We thereby conclude
that the statistical gauge-field excitations
remain gapless at all temperatures in the present AH$^2$ model.

{\it A. Low-Temperature Phase.}  Let us first
 consider the Wilson loop for a large contour
 $C$ in low temperature limit.  As in the previous case of the phase
auto-correlation functions, it can be expressed as
$$W(C)=
\Biggl\langle {\rm exp}
\Biggl(ie^{\prime}\oint_C A_{\mu}dr_{\mu}\Biggr)
\Biggr\rangle
 = {Z^{\prime}\over Z}, \eqno (46)$$
where the partition function $Z^{\prime}$ differs
 from $Z$
only by the addition of the exponent
$-ie^{\prime}\oint_C A_{\mu}dr_{\mu}$
to the energy functional (4).  Again, following
refs. 18 and 19, $Z^{\prime}$
may be computed via the Villain
duality transformation by substitution of the
 mathematical identity (12), which is valid
in the low-temperature limit.  Generalizing the
treatment of Jones et al.,$^{11}$ we find
 ultimately that $Z^{\prime}$ has the following ``roughening model''
representation:
$$\eqalignno{Z^{\prime}=\sum_{\{n_b(r)\}}
\sum_{\{n_f(r)\}}
{\rm exp}\Biggl\{ & -{1\over{2\beta_b}}
\sum_{r,\mu}[\Delta_{\mu}n_b(r)]^2
-{1\over{2\beta_f}}\sum_{r,\mu}
[\Delta_{\mu}n_f(r)]^2\cr
&-{(qg)^2\over 2}\sum_r\Bigl[n_b(r)+n_f(r)+
{e^{\prime}\over{qg}}J(r)\Bigr]^2
\Biggr\},& (47)\cr}$$
where
$J(r)$ has a value of
$1$ if the point $r$ is within the contour $C$,
and it vanishes otherwize.
Again, after the application of the Poisson summation
formula, and subsequently integrating over the fields
 that this formula generates, we find
that $Z^{\prime}$ is given by the product a trivial
 ``spin-wave'' factor with the following Coulomb
gas enesemble average:
$$\eqalignno{Z_{\rm Coulomb}^{\prime}=
&\sum_{\{q_{\alpha}(r)\}}{\rm exp}\Biggl\{-(2\pi)^2\bar\beta {1\over
2}\sum_{r,r^{\prime}}\Biggl[q_{\alpha}(r)
G_{\alpha\beta}(r,r^{\prime})q_{\beta}(r^{\prime})
+i\pi^{-1}e^{\prime}qg q_{\alpha}(r)G_{\alpha\beta}^{({\rm
sr})}(r,r^{\prime})u_{\beta}J(r^{\prime})\cr
&-(2\pi)^{-2}e^{\prime 2}(qg)^2 J(r)
u_{\alpha}G_{\alpha\beta}^{({\rm sr})}
(r,r^{\prime})u_{\beta}J(r^{\prime})
+(2\pi)^{-2}e^{\prime 2}\bar\beta^{-1}
J(r)\delta_{r,r^{\prime}}\Biggr]\Biggr\},& (48)\cr}$$
again where
$\vec u=(1,1)$ represents the spin/charge
components of the statistical degree of
freedom (25), and where $G_{\alpha\beta}^{({\rm sr})}
(r,r^{\prime})$ represents the short-range component
 of the matrix Greens function [see Eqs. (16-19)].
The above  expression constitutes the Coulomb gas
representation of the modified partition function $Z^{\prime}$.

To further reduce expression (48) for
 $Z_{\rm Coulomb}^{\prime}$, we recall that
$$\sum_r G_{\alpha\beta}^{({\rm
sr})}(r,0)={\beta_\alpha\beta_\beta\over{\beta_b\beta_f}}
\lambda_{\rm st}^2.$$
Hence, in the limit of large contours, $C$, the
sum over the relative coordinates in the
third term of this expression can be extended
 to the entire lattice.  In this case we find
 that the last two terms in (48) sum to zero; i.e.,
$$(2\pi)^{-2}e^{\prime 2}
\Biggl[-\sum_{r}{(qg)^2\over{\beta_b\beta_f}}
(\beta_b+\beta_f)^2\lambda_{\rm st}^2J(r)+
\sum_{r}\Biggl({1\over{\beta_b}}
+{1\over{\beta_f}}\Biggr)J(r)\Biggr] = 0,$$
by (20).  Note that the first term above
corresponds to the third term in (48),
where the sum over the relative coordinates
 has been already performed.
Similarly, since $G_{\alpha\beta}^{({\rm sr})}
(r,r^{\prime})$ is short-ranged,
we have that in the limit of large contours,
$$\sum_{r,r^{\prime}}q_{\alpha}(r)G_{\alpha\beta}^{({\rm
sr})}(r,r^{\prime})u_{\beta}J(r^{\prime})
={\beta_b+\beta_f\over{\beta_b\beta_f}}\lambda_{\rm st}^2
\sum_r[\beta_b q_b(r)+\beta_f q_f(r)]J(r).$$
The lefthand side of this expression is proportional
to the second term in (48),
which implies that the Wilson
loop, $Z^{\prime}/Z$, is given by
$$W(C)=\Biggl\langle{\rm exp}
\Biggl[-2\pi i e^{\prime}qg
\lambda_{\rm st}^2\sum_r[\beta_b q_b(r)+\beta_f
q_f(r)]J(r)\Biggr]\Biggr\rangle_{\rm Coulomb}.\eqno (49)$$
Since vortices are bound to anti-vortices
 at low temperature due to the existence
of long-range interactions, we may employ
a cummulant expansion
to compute this average, as in the previous
calculation of phase correlators [see Eqs. (36) and (37)].  Such an
approximation yields
$$\eqalignno{W(C)&={\rm exp}
\Biggl\{-(2\pi e^{\prime}qg)^2
\lambda_{\rm st}^4{1\over 2}
\sum_{r^{\prime},r^{\prime\prime}}
\langle [\beta_b q_b(r^{\prime})+
\beta_f q_f(r^{\prime})]
[\beta_b q_b(r^{\prime\prime})+
\beta_f q_f(r^{\prime\prime})]\rangle
J(r^{\prime})J(r^{\prime\prime})\Biggr\}\cr
&={\rm exp}\Biggl\{-\Biggl({2\pi
e^{\prime}\over{qg}}\Biggr)^2\Biggl[1+{\rm
max}\Biggl({\beta_b\over\beta_f},{\beta_f\over\beta_b}\Biggr)\Biggr]^{-2}
{1\over 2}\sum_{r^{\prime},r^{\prime\prime}}\langle
q_{i}(r^{\prime})q_{i}(r^{\prime\prime})\rangle
J(r^{\prime})J(r^{\prime\prime})\Biggr\}, & (50)\cr}$$
again where the species label $i$ is such that $\beta_{i}={\rm
min}(\beta_b,\beta_f)$.
Since the charge correlations decrease rapidly,
we may expand $J(r^{\prime})$ and $J(r^{\prime\prime})$
 with reference to the center of
mass coordinates; i.e.,
$$\eqalignno{J(r^{\prime}) =&J(R)+
{1\over 2}\vec r\cdot\vec\nabla J(R),\cr
    J(r^{\prime\prime}) =&J(R)-
{1\over 2}\vec r\cdot\vec\nabla J(R),\cr}$$
where $R$ is the center of mass coordinate
and $r$ is the relative coordinate between
points $r^{\prime}$ and $r^{\prime\prime}$.
  Due to  the charge neutrality condition
 effectively imposed by the presence of
long-range interactions, only the last terms
above remain in sum over coordinates found in
 the exponent of expression (50).  This
ultimately results in
$$\eqalignno{W(C)&={\rm exp}\Biggl\{{1\over 2}\Biggl({2\pi
e^{\prime}\over{qg}}\Biggr)^2\Biggl[1+{\rm
max}\Biggl({\beta_b\over\beta_f},{\beta_f\over\beta_b}\Biggr)\Biggr]^{-2}
\sum_{r}{1\over 2}r^2\langle q_{i}(0)q_{i}(r)\rangle \sum_R{1\over 4}
[\vec\nabla J(R)]^2\Biggr\}\cr
&={\rm exp}\Biggl\{{\pi^2\over 4} \Biggl({e^{\prime}\over{qg}}\Biggr)^2
\Biggl[1+{\rm max}
\Biggl({\beta_b\over\beta_f},{\beta_f\over\beta_b}\Biggr)
\Biggr]^{-2}\Biggl[\sum_{r}r^2\langle q_{i}(0)q_{i}(r)
\rangle\Biggr]P\Biggr\}, & (51)\cr}$$
where that prefactor of ${1\over 2}$ in the sum
over the relative coordinates arises
from an angular average.  Above, $P$ denotes the
 length of the contour $C$.  The latter expression
 represents the final result of our manipulations.

We observe, therefore,  that at the low temperatures
 presently considered, the Wilson loop exhibits a
 perimeter law.
And since the vortex charge correlator,
$\langle q_{i}(0)q_{i}(r)\rangle$, corresponding
to the species with the least costly vortex
 excitations is given by Eq. (40),
we also observe
that the coefficient to the above perimeter-law
diverges precisely at the BKT phase transition
temperature (9).  This suggests that a ``confining''
 area law behavior exists above $T_c$, which
we in fact argue for below.  Last,
it is also of interest to remark that this
coefficient vanishes in the limit $J_b\gg J_f$ or $J_f\gg J_b$.

{\it B. High-Temperature Phase.}  Clearly,
 in the high-temperature limit,
$\beta_{b}=0$ and $\beta_{f}=0$, the
present AH$^2$ reduces to a pure Abelian gauge theory,
which is trivially ``confining'' in 1+1
 dimensions, and which therefore shows an area law for
the Wilson loop.  Also, in either the
 limits $\beta_b=0$ or $\beta_f=0$
that correspond to the one-component model,
it is well know that the Wilson loop
follows an area law at all temperatures
 in the case of fractional probe
 charges $e^{\prime}$.$^{11}$
 It has recently been argued, however,
that such ``confinement'' is trivially
due to pure 1+1 dimensional
statistical ``electro-magnetism''
that is renormalized ``dielectrically''.$^{21}$
In this case, one finds that
 gauge-field fluctuations
are characterized by a free $U(1)$
action [see the last term in Eq. (4)],
 with a renormalized ``charge'',
$g/\epsilon_{\rm st}^{1/2}$, that is on the order
of $g$ for temperatures above a cross-over
 temperature scale,$^8$ $T_{\rm c/o}
\lsim J_s\ (s=b\,{\rm or}\, f)$,
and that becomes exponentially
small below this cross-over
temperature scale.$^{21}$  This ultimately
leads to a cross-over phenomenon
between a ``strange'' metal phase at high
temperatures and its absence at low
temperature.$^{8}$
Given that the only true phase transition
 in the present AH$^2$
model is the BKT transition itself, by continuity
in the phase diagram (see Fig. 1), ({\it i}) the
Wilson loop should follow a ``confining'' area
law at all temperatures above $T_c$ for fractional probe
 charges $e^{\prime}$ and ({\it ii})
we expect that a similar crossover phenomon
occurs in the present AH$^2$ model
in this temperature regime,
with the exception that the crossover
temperature scale is given by
$T_{\rm c/o}
\lsim {\rm max}(J_b,J_f)$ in this case.

In summary, the Wilson loop (46) of the
statistical gauge-field in the presently
considered AH$^2$ model shows a perimeter law at
low temperatures $T<T_c$, such that the corresponding coefficient
 diverges at $T_c$.  Also, in the general case of fractional
 probe charges,
 $e^{\prime}\neq nq$ ($n=0, \pm 1,\pm 2, ...$), the
 Wilson loop shows an area law for temperatures
above the BKT transition, which we argue implies a
gapless spectrum for statistical gauge-field excitations.
 Since we showed in the previous section that there exists
 no Higgs mechanism for the statistical gauge-field to
acquire a gap due to the complete absence of long-range
order in the statistical phase (25), and since the perimeter
 law exhibited by the Wilson loop in the low temperature
 phase does not indicate the existence of short-range
``electromagnetic'' interactions, we argue that the
statistical gauge-field remains gapless as well
 in this phase.  The occurence of a perimeter
 law, therefore, is simply a signal of the
vortex binding-unbinding transition present in the system.

\bigskip
\bigskip
\centerline{\bf VI. Universality  Class}
\bigskip
By consideration of both the continuum limit
 (section III) and the low-temperature limit
(section IV), we have shown above that a short-range
order to algebraic long-range order phase transition
occurs at a transition temperature given by Eq. (9)
for the physical electronic phase (8) in the present
 AH$^2$ model for spin-charge separated superconductivity.
But what is the nature of this transition in physical terms?
Below, we argue that the fore-mentioned transition is
 in the same universality
class as that of the XY model.  Hence, we expect a
corresponding universal jump in the superfluid density.$^{16}$

Let us now derive the renormalization group equations
for the BKT transition in the physical electronic phase
of the present model by closely following the treatment
 of the corresponding problem in the case of the XY model.$^{19}$
{}From eq. (44), we have that
$${1\over{\beta_{\rm eff}}}={1\over{\bar\beta}}+
\pi^3y^2\int_a^{\infty}{dr\over a}\Biggl({a\over
r}\Biggr)^{2\pi\tilde\beta(r)-3},\eqno (52)$$
where by Eq. (40) for the charge correlations,
$\tilde\beta(r)$ is a smooth function satisfying
$$\eqalignno{&\beta_i; \quad\ r\ll \lambda_{\rm st} & (53a)\cr
\tilde\beta(r)=\qquad\cr
             &\bar\beta; \quad r\gg\lambda_{\rm st} & (53b)\cr}$$
and where the activity at $r=r_0$ is given by
$$y={\rm exp}\Biggl[-\pi\tilde\beta(r_0)\,{\rm ln}
{a\over{r_0}}\Biggr].\eqno (54)$$
Let $x=r/a$, and suppose that we make the following
approximation to the integral appearing in the
rigthhand side of Eq. (52):
$$\eqalignno{
\int_1^{\infty}dx x^{3-2\pi\tilde\beta(x)}&\cong\int_1^{\lambda_{\rm
st}}dxx^{3-2\pi\beta_i}+
\int_{\lambda_{\rm st}}^{\infty}dx x^{3-2\pi\bar\beta}\cr
&={\lambda_{\rm st}^{4-2\pi\beta_i}-1\over{4-2\pi\beta_i}}+{\lambda_{\rm
st}}^{4-2\pi\bar\beta}\int_1^{\infty}dx x^{3-2\pi\bar\beta}\cr
&\rightarrow {\rm ln}\,\lambda_{\rm st}+[1+(4-2\pi\bar\beta){\rm
ln}\,\lambda_{\rm st}]\int_1^{\infty}dx x^{3-2\pi\bar\beta}& (55)\cr}$$
in the limit $\lambda_{\rm st}\rightarrow 1$.
Hence, in this limit, we have by (52) that
$${1\over{\beta_{\rm eff}}}={1\over{\bar\beta}}+
\pi^3y^2{\rm ln}\,\lambda_{\rm st}+
\pi^3y^2[1+(4-2\pi\bar\beta){\rm ln}\,\lambda_{\rm st}]\int_1^{\infty}dx
x^{3-2\pi\bar\beta}.\eqno (56)$$
This formula may be rexpressed as
$${1\over{\beta_{\rm eff}}}={1\over{\beta_{\lambda}}}+
\pi^3y_{\lambda}^2\int_1^{\infty}dx x^{3-2\pi\bar\beta},
\eqno (57)$$
where
$$\eqalignno{
{1\over{\beta_{\lambda}}}&={1\over{\bar\beta}}+
\pi^3y^2{\rm ln}\,\lambda_{\rm st}, & (58)\cr
y_{\lambda}&=y+(2-\pi\beta_{\lambda})y\,{\rm ln}\,\lambda_{\rm st}.
 & (59)\cr}$$
The differential form of the latter two relations are simply
the Kosterlitz renormalization group equations$^{16,19}$
$$\eqalignno{
\lambda_{\rm st}{d\over{d\lambda_{\rm
st}}}\beta_{\lambda}^{-1}&=\pi^3y_{\lambda}^2, & (60)\cr
\lambda_{\rm st}{d\over{d\lambda_{\rm
st}}}y_{\lambda}&=(2-\pi\beta_{\lambda})y_{\lambda},& (61)\cr}$$
with the length  renormalization scale determined by the model parameter
$\lambda_{\rm st}$.

In summary,  given the validity of
approximation (55), we find the BKT
 transition obtained previously in
the present AH$^2$ model for the
physical electronic phase is in
the same universality class as
that of the XY model.  We therefore
expect a universal jump in the
superfluid density, following that predicted for the XY model.$^{16}$

\bigskip
\bigskip
\centerline{\bf VII. Summary and Discussion}
\bigskip

In this paper, we have introduced
a square lattice gauge theory model
for spin-charge separated superconductivity
in the presence of statistical gauge-field excitations.
 Such excitations appear naturally in spin-charge
 separated descriptions of strongly interacting
electrons systems, and they represent so-called
 chiral spin fluctuations.$^{10}$  After analysing
this model in various limits (continuum and low-temperature),
we found that it explicitely shows a
superconducting transition of the
BKT type for the physical electronic phase (8)
 at a critical temperature given by (9).
 As expected, we see that statistical
gauge field excitations suppress the
 superfluid transition temperature
with respect to mean-field; i.e.,
$T_c<T_c^{(0)}$. It was also argued
in the preceeding section that this transition shares the same
universality class as that of the XY model, and hence that
we expect a
corresponding universal jump in the superfluid density.
All other phases, however,  were found to show only
 short-range auto-correlations
at low-temperature.  In particular, the fact that statistical
phase (25) correlations are short-range at
 all temperatures implies that
the corresponding transverse statistical
 gauge-field fluctuations
do not acquire a gap via the Higgs mechanism.
Furthermore, we have argued that because
of dimensionality, neither
``confinement'' effects produce a gap
 in these excitations.  Hence, we claim that the
statistical gauge-field excitations remain gapless
at all temperature in the present model.  It is
important to note, however,  that gapless statistical
 gauge-field excitations also have been shown to exist
 at the meanfield level in the commensurate flux phase
 of the $t-J$ model, which is thought to be a spin-charge
 separated anyonic superconductor, but they were to
found acquire a gap once dynamical effects were included.$^{22}$
It is therefore not certain that such gapless excitations
 will persist in
the present model once such effects are accounted for.
 We hope to address this issue in a future publication.

The superfluid transition found in the AH$^2$ model
 for spin-charge separated superconductivity can be understood {\it a
posteriori} as follows.  Let
$\Psi_f=\langle c_{i\uparrow}c_{i^{\prime}\downarrow}\rangle$ and
$\Psi_b=\langle b_i b_{i^{\prime}}\rangle$ be
the respective order parameters of each specie.
 Now suppose  that the superconductivity in this
 spin-charge separated system results from
conventional Cooper pairing of electrons, with
an order parameter given by $\Psi_{\rm el}=\langle
c_{i\uparrow}b_i^{\dag}c_{i^{\prime}\downarrow}
b_{i^{\prime}}^{\dag}\rangle$.  Then the mean-field
approximation yields that the true superconducting
 order parameter satisfies
$$\Psi_{\rm el}\cong\Psi_f\Psi_b^*\propto
 e^{i(\phi_f-\phi_b)}.$$
It is thus not surprising to find that the
physical electronic phase (8)
develops quasi-long-range order in the present model.
 What is indeed important to note, however, is that fluctuations in the
statistical gauge field do not suppress the transition altogether.

Yet can the present model explain, or at least describe,
 the phenomenon of high-temperature superconductivity?
 Mean-field RVB treatments of the $t-J$ model yield a
superfluid phase stiffness for spinon pairs of
$J_f\sim J(1-x/x_0)$, where $x$
denotes the hole concentration and $x_0$ denotes the
critical concentration beyond which there is no pairing
instability.$^9$  Also, since we have assumed throughout
 that the holons form pairs, then their corresponding
superfluid phase stiffness is  given by $J_b\sim t^{\prime}x$, where
$t^{\prime}$ denotes the hopping matrix element for such
pairs.  Now suppose  that the latter hopping matrix element satisfies
 $t^{\prime}\sim 1000\,{\rm K}$, which is
conceivable since the corresponding matrix elements
for a single electron is on the order of $t\sim 5000\,{\rm K}$.
Then since  Eq. (9) implies that the
superconducting temperature satisfies
$T_c\lsim {\rm min}(t^{\prime}x_0, J)$,
and since  $J\sim 1000\,{\rm K}$ while
 the maximum hole concentrations for oxide superconductivity
are typically near $x_0\cong 0.2$,  we have that
$T_c\lsim t^{\prime}x_0\sim 200\,{\rm K}$, in agreement
 with experiment.$^1$   More importantly, however,
Eq. (9) for the critical temperature coupled with the
above functional dependences for the superfluid
stiffnesses of each species with hole concentration  imply  a
phase-diagram for the superconducting versus normal
 phase that is shown in Fig. 1.  The shape of the
phase boundary qualitatively resembles that of the high-temperature
superconductors,$^{20}$  but it differs
substantially in shape from that predicted
by Ginzburg-Landau treatments of spin-charge separated
superconductivity.$^8$
Finally, since
$g^2=T/\chi_d$ in this case,
where $\chi_d=\chi_f+\chi_b\sim J(1-x)+tT_b/T$
is the sum of the diamagnetic
susceptibilities of each species,$^{5,6}$
  with $T_b\lsim tx$, we then have that
$\lambda_{\rm st}=
q^{-1}[\chi_d/(J_b+J_f)]^{1/2}\gsim 1$.  Hence, the
 renormalization group analysis discussed in the previous
section is valid, and we therefore expect
a universal jump in the superfluid density at the critical
temperature in the absence of interplanar coupling.  Note
 that the present description of oxide superconductivity
differs substantially from that proposed by Anderson
and coworkers,$^{28}$ where it is assumed that interlayer
interactions drive the phenomenon.  Although it is
proposed here, on the contrary, that oxide superconductivity
is primarily a one-layer effect, we nevertheless hope to
study the problem of a few coupled layers of square lattice
 AH$^2$ models in a future publication.  It is quite possible
that the nature of the presently discovered superfluid transition
changes dramatically
once the model is extended into the third dimension,
which could be relevant to the layered
high-temperature superconductors.$^8$

Finally, what is the nature of the normal state above $T_c$?
 It was previously remarked in section II that
separate ``strange'' metal, Fermi-liquid, superconducting,
 and spin-gap phases
exist in certain spin-charge separated treatments of the
$t-J$ model in two dimensions at the mean-field level,$^8$
 as shown by the dashed lines in Fig. 1.
We remind the reader that the ``strange'' metal phase
corresponds to the absence of superfluidity in both species,
the Fermi-liquid phase to the appearence of superfluidity
in the holon species alone, the superconducting phase to
the appearance of superfluidity in both species, and the
spin-gap phase to the existence of superfuidity in the
spinon species alone.
In this paper, we have demonstrated explicitely that the
only true phase transition that remains beyond the meanfield
approximation is the superconduting one depicted by the solid
line in Fig. 1.  In particular, fluctuations in the statistical
 gauge field representing chiral spin-fluctuations$^{10}$
destroy the meanfield transitions between the ``strange''
metal phase and the Fermi-liquid phase, and between the
``strange'' metal phase and the spin-gap phase.$^8$
This effect can be studied more easily by considering
 the case when only one of the species has a superfluid
instability; i.e., the one-component Abelian-Higgs
model corresponding to $\beta_b=0$ or $\beta_f=0$.
  It has been explicitely shown in this case, by using
techniques similar to those employed in this paper,
 that the BKT superfluid transition present in the
absence of statistical gauge-field fluctuations
 becomes a cross-over phenomenon.$^{21}$  The
 cross-over temperature scale satisfies
$T_{\rm c/o}\lsim T_s$, where $T_s$ denotes
the meanfield transition temperature of the
species in question, and it {\it vanishes}
in the limit of maximum gauge-field fluctuations,
 $\lambda_{\rm st}\rightarrow 0$.  By continuity
with the present AH$^2$ model, we expect that this
 crossover phenomenon persists in the normal phase
above $T_c$; i.e., $T_{\rm c/o}\lsim {\rm max}(T_b, T_f)$.
  Furthermore, in the limit $\lambda_{\rm st}\rightarrow 0$,
 the normal phase should be entirely ``strange'',
with a characteristic $T$-linear resistivity.
 As discussed above, however, reasonable
 estimates based on the $t-J$ model yield a
statistical length scale satisfying $\lambda_{\rm st}\gsim 1$.
Hence, we expect that the cross-over phenomenon
 from the ``strange'' metal regime into the
 Fermi-liquid and spin-gap regimes should
 persist in strongly interacting 2D electron systems.
Such a cross-over scale
(see the upper lines in Fig. 1)
is consistent with the narrow
doping window  for linear-in-$T$
resistivity recently observed
in the superconducting oxides.$^{29}$

The author gratefully acknowledges
illuminating discussions
with S. Trugman, P. Lederer, D. Callaway,
J. Kogut, J. Palmeri, E. Fradkin and M. Inui.
This work was supported in
part by NATO grant CRG-920088.

\vfill\eject
\centerline{\bf Appendix A: ``Spin-Wave''
Component of Phase Correlator}
\bigskip
To calculate the ``spin-wave'' component
of the phase auto-correlation
function (31) in the AH$^2$ model at low-temperature, given by
$$G_{\rm wave}(\vec r_{12})={\rm exp}\,X_{\rm wave},\eqno (A.1)$$
 with
$$\eqalignno{X_{\rm wave}=&-{1\over 2}\Biggl({p_b^2\over{\beta_b}}
+{p_f^2\over{\beta_f}}\Biggr)\sum_{r,\mu}\eta_{\mu}^2(r)\cr
&-{1\over{\pi}}{\bar\beta\over
{\beta_b\beta_f}}\sum_{r,r^{\prime}}
\eta(r)[p_{\rm el}^{\prime 2}
\Gamma_{\rm lr}(\vec r-\vec r\,^{\prime})
+p_{\rm st}^2\Gamma_{\rm sr}
(\vec r-\vec r\,^{\prime})]\eta(r^{\prime}), & (A.2)\cr}$$
 we will follow the standard treatment of the
 corresponding problem for the XY model.$^{18,19}$
Let us suppose that $\vec r_{12}=\vec r_1-\vec r_2$
is directed along the
x-axis.  Then clearly,
$$\eqalignno{X_{\rm wave}=&-{1\over 2}
\Biggl({p_b^2\over{\beta_b}}
+{p_f^2\over{\beta_f}}\Biggr)r_{12}\cr
&-{1\over{\pi}}{\bar\beta\over
{\beta_b\beta_f}}\sum_{x,x^{\prime}=0}^{r_{12}-1}
p_{\rm el}^{\prime 2}
[2\Gamma_{\rm lr}(x-x^{\prime},0)
-\Gamma_{\rm lr}(x-x^{\prime},1)
-\Gamma_{\rm lr}(x-x^{\prime},-1)
] \cr
&-{1\over{\pi}}{\bar\beta\over
{\beta_b\beta_f}}\sum_{x,x^{\prime}=0}^{r_{12}-1}
p_{\rm st}^2[2\Gamma_{\rm sr}
(x-x^{\prime},0)
-\Gamma_{\rm sr}
(x-x^{\prime},1)
-\Gamma_{\rm sr}(x-x^{\prime},-1)]. & (A.3)\cr
}$$
But since $\Gamma_{\rm lr}(\vec r)$ and
$\Gamma_{\rm sr}(\vec r)-\Gamma_{\rm sr}(\infty)$
are proportional to the respective long-range and
short-range lattice Greens functions, then
by definition [see Eqs. (16), (18) and (19)] they satisfy
$$\eqalignno{
4\Gamma_{\rm lr}(x,y)-
[\Gamma_{\rm lr}(x+1,y)+\Gamma_{\rm lr}(x-1,y)
\ \ \ \ \ \ \ \ \ \ \ \ \ \ \ \ \ \ \ \ \ \ \ &\cr
+\Gamma_{\rm lr}(x,y+1)
+\Gamma_{\rm lr}(x,y-1)]&=-2\pi\delta_{x,0}\delta_{y,0},& (A.4)\cr
\cr
4\Gamma_{\rm sr}(x,y)-
[\Gamma_{\rm sr}(x+1,y)+\Gamma_{\rm sr}(x-1,y)
\ \ \ \ \ \ \ \ \ \ \ \ \ \ \ \ \ \ \ \ \ \ \ &\cr
+\Gamma_{\rm sr}(x,y+1)
+\Gamma_{\rm sr}(x,y-1)]+\lambda_{\rm st}^{-2}[\Gamma_{\rm sr}(x,y)
-\Gamma_{\rm sr}(\infty)]
&=-2\pi\delta_{x,0}\delta_{y,0}.
& (A.5)\cr}$$
Substituting in the latter identities into
(A.3) for the Greens function terms at points
off of the $y=0$ axis, and then using the fact
 that the resulting sum over $x$ and $x^{\prime}$
 is a partially telescoping series,$^{18,19}$
we find that
$$\eqalignno{X_{\rm wave}=&-{1\over 2}\Biggl
({p_b^2\over{\beta_b}}
+{p_f^2\over{\beta_f}}\Biggr)r_{12}-
{1\over{\pi}}{\bar\beta\over
{\beta_b\beta_f}}
p_{\rm el}^{\prime 2}
[2\Gamma_{\rm lr}(r_{12})
-2\pi r_{12}] \cr
&-{1\over{\pi}}{\bar\beta\over
{\beta_b\beta_f}}p_{\rm st}^2
\Biggl\{2\Gamma_{\rm sr}(r_{12})
-2\pi r_{12}-\lambda_{\rm st}^{-2}
\sum_{x,x^{\prime}=0}^{r_{12}-1}[\Gamma_{\rm sr}(x-x^{\prime})
-\Gamma_{\rm sr}(\infty)]\Biggr\},\cr
=&-{2\bar\beta\over
{\beta_b\beta_f}}
p_{\rm el}^{\prime 2}
{\Gamma_{\rm lr}(r_{12})\over{\pi}}\cr
&-{2\bar\beta\over
{\beta_b\beta_f}}p_{\rm st}^2
\Biggl\{{\Gamma_{\rm sr}(r_{12})\over{\pi}}
+\lambda_{\rm st}^{-2}\sum_{x,x^{\prime}=0}^{r_{12}-1}
{1\over{2\pi}}[\Gamma_{\rm sr}(\infty)-\Gamma_{\rm sr}(x-x^{\prime})
]\Biggr\}.& (A.6)\cr
}$$
But in the limit $r_{12}\gg\lambda_{\rm st}$, we have the identity
$$\lambda_{\rm
st}^{-2}\sum_{x,x^{\prime}=0}^{r_{12}-1}{1\over{2\pi}}[\Gamma_{\rm
sr}(\infty)-\Gamma_{\rm sr}(x-x^{\prime})]\cong \lambda_{\rm
st}^{-2}\sum_{x=0}^{r_{12}-1}
\sum_{x^{\prime}=0}^{\infty}
{1\over{2\pi}}[\Gamma_{\rm sr}(\infty)-
\Gamma_{\rm sr}(x-x^{\prime})]=r_{12}.$$
Substituting this relationship into the
last term of (A.6), we find by (A.1)
 that the ``spin-wave'' component of
the phase correlator in the AH$^2$
model on the square-lattice is given by
$$G_{\rm wave}(\vec r_{12})=
{\rm exp}\Biggl[-{2\bar\beta\over{\beta_b\beta_f}}
p_{\rm st}^2(r_{12}+\pi^{-1} l{\rm n}\,\lambda_{\rm st})\Biggr]
{\rm exp} \Biggl[-{2\bar\beta\over{\beta_b\beta_f}}
p_{\rm el}^{\prime 2}{\Gamma_{\rm lr}(r_{12})
\over{\pi}}\Biggr]\eqno (A.7)$$
in the limit $|\vec r|\gg\lambda_{\rm st}$, where
$$l{\rm n}\,\lambda_{\rm st}=\Gamma_{\rm sr}(\infty)=(2\pi)^{-1}
\int_{\rm BZ}d^2k
(\tilde k^2+\lambda_{\rm st}^{-2})^{-1}.\eqno (A.8)$$
Note, of course, that the function defined
 above coincides with the Napierian logarithm in the limit
$\lambda_{\rm st}\gg 1$.

\bigskip
\bigskip
\centerline{\bf Appendix B: Phase Correlations in One-component Model}
\bigskip
Consider the gauge-invariant phase correlation
function
$$C_{12}=\langle {\rm exp}\{ip_b[\phi_b(r_1)-
\phi_b(r_2)
+q\int_{r_1}^{r_2}A_{\mu}dr_{\mu}]\}\rangle = Z^{\prime}/Z\eqno (B.1)$$
in the case of the one-component Abelian-Higgs model $\beta_f=0$,
where the partition function $Z^{\prime}$ differs from $Z$
only by the addition of the exponent $-ip_b[\phi_b(r_1)-
\phi_b(r_2)+q\int_{r_1}^{r_2}A_{\mu}dr_{\mu}]$
to the energy functional (4). Then by (27),
we have that the Villain
duality transformation of this partion function is given by
$$Z^{\prime}=\sum_{\{n_b(r)\}}{\rm exp}\Bigl\{-{1\over{2\beta_b}}
\sum_{r,\mu}[\Delta_{\mu}n_b(r)+p_b\eta_{\mu}(r)]^2
-{g^2q^2\over 2}\sum_r n_b^2(r)\Bigr\}.\eqno (B.2)$$
As discussed in section IV, $n_b(r)$ ranges over the
 integers and $r$ covers the dual
lattice,
while the dual lattice ``dipole'' vector,
$\eta_{\mu}(r)$, has a value of
$\pm 1$ if it intersects a fixed path
on the original lattice connecting
points $\vec r_1$ and $\vec r_2$,
and it vanishes otherwize.$^{18,19}$  Therefore,
in the limit $\lambda_{\rm st}\ll 1$,
where statistical gauge-field fluctuations are maximized,
the trivial configuration $n_b(r)=0$ dominates the
 ensemble average above,
resulting in
$$C_{12}\cong
{\rm exp}\Bigl(-{p_b^2\over{2\beta_b}}
|\vec r_1-\vec r_2|\Bigr)
.\eqno (B.3)$$
Notice, as expected [see Eq. (9)], that this
correlation function shows no  evidence for a
 phase-transition at non-zero temperature.

It was also shown previously that the
 application of the Poisson summation formula to the ``roughening model''
representation
expression (B.2) leads ultimately to the factorization
 $C_{12} \cong G_{\rm wave}(\vec r_1-\vec r_2)
G_{\rm Coulomb}(\vec r_1-\vec r_2)$ of
 the gauge-invariant phase correlation function.
 In the present case of $\beta_f=0$ and $p_f=0$,
Eq. (A.7) yields that
$$G_{\rm wave}(\vec r_1-\vec r_2)\cong
{\rm exp}\Bigl[-{p_b^2\over{2\beta_b}}
(|\vec r_1-\vec r_2|+\pi^{-1} l{\rm n}\, \lambda_{\rm st})\Bigr]
\eqno (B.4)$$
for $r\gg\lambda_{\rm st}$, while Eq. (36) yields that
$$G_{\rm Coulomb}
(\vec r_1-\vec r_2)=
\Bigl\langle{\rm exp}\Bigl\{i
p_b\sum_{r}
q_b(r)
\sigma_{\rm sr}(r)
\Bigr\}\Bigr\rangle_{\rm Coulomb}, \eqno (B.5)$$
 where  the corresponding partition function (21) for
the screened Coulomb gas ensemble is given by$^{25}$
$$Z_{\rm Coulomb}=\sum_{\{q_b(r)\}}{\rm
exp}\Biggl\{2\pi\beta_b\sum_{(r,r^{\prime})}
\Gamma_{\rm sr}(\vec r-\vec r\,^{\prime})
q_{b}(r)q_{b}(r^{\prime})]\Biggr\}.
\eqno (B.6)$$
Unlike the previous treatment in the case of the AH$^2$ model,
however, we now take the original definition
$$\Gamma_{\rm sr}
(\vec r)=-\int_{\rm BZ}{d^2k\over{2\pi}}
e^{i\vec k\cdot\vec r}
{1\over{\tilde k^2+\lambda_{\rm st}^{-2}}}, \eqno (B.7)$$
for the interaction energy, since charge conservation is no
longer guaranteed due to the absence of long-range forces.
Following ref. 11,
in the limit $\lambda_{\rm st}\gg 1$, in which case the
interaction between
vortices is negligible at large distances, we may
approximate the above enemble average by
$$Z_{\rm Coulomb}\cong\Pi_r\sum_{q_b(r)=-\infty}^{\infty}{\rm
exp}\Biggl[-2\pi\beta_b {1\over 2}(l{\rm n}\, \lambda_{\rm st})
q_{b}^2(r)\Biggr].
\eqno (B.8)$$
Hence, if we truncate the sum over vortex charge
to $q_{b}(r)=0,\pm 1$, then by (B.5),
$$\eqalignno{G_{\rm Coulomb}
&\cong{\Pi_r\{1+2\,{\rm exp}(-\pi\beta_b l{\rm n}\,
\lambda_{\rm st}){\rm cos}\,
 [p_b\sigma_{\rm sr}(r)]\}\over{\Pi_r[1+2\,{\rm exp}(-\pi\beta_b\,
l{\rm n}\, \lambda_{\rm st})]}}\cr
&\cong {\rm exp}(2e^{-\pi\beta_b\, l{\rm n}\,
\lambda_{\rm st}}\sum_r\{{\rm cos}\,
[p_b\sigma_{\rm sr}(r)]-1\})\cr
&\cong {\rm exp}(2\lambda_{\rm st}^{-\pi\beta_b} \sum_r\{{\rm cos}\,
[p_b\sigma_{\rm sr}(r)]-1\})\cr
&\rightarrow 1\cr}$$
in the limit $\lambda_{\rm st}\rightarrow \infty$ and
$\beta_b\rightarrow\infty$.
Thus, in these limits, the gauge-invariant phase
correlation function, $C_{12}$, is approximately
given by the ``spin-wave'' contribution (B.4) for $r\gg\lambda_{\rm st}$.

In the low-temperature limit of the one-component Abelian-Higgs model,
therefore,
we generally find that the coherence length
for the holon phase correlations is given by
$\xi_b\sim\beta_b=J_b/T$ in units of the lattice constant,
$a$, contrary to previous claims in the
literature of an exponentially divergent length.$^{8}$

\vfill\eject
\centerline{\bf References}
\vskip 16 pt

\item {1.} {\it The Physical Properties
of High-Temperature Superconductors},
vol. 1, edited
by D.M. Ginsberg (World Scientific, Singapore, 1989).

\item {2.}  P.W. Anderson, Science
 {\bf 235}, 1196 (1987); see also "Frontiers and Borderlines in
Many-body Physics", Varenna Lectures (North Holland,
Amsterdam, 1987).

\item {3.} L.B. Ioffe and A.I. Larkin,
 Phys. Rev. B {\bf 39}, 8988 (1989).

\item {4.} J.P. Rodriguez and
 B. Dou{\c c}ot, Europhys. Lett.
{\bf 11}, 451 (1990).

\item {5.} N. Nagaosa and P.A.Lee,
Phys. Rev. Lett. {\bf 64},
2450, (1990);  L.B. Ioffe and
 P.B. Wiegmann, Phys. Rev. Lett. {\bf 65},
653 (1990); L.B. Ioffe and
 G. Kotliar, Phys. Rev. B {\bf 42}, 10348 (1990).

\item {6.} J.P. Rodriguez, Phys. Rev.
 B {\bf 44}, 9582 (1991); (E)
{\bf 45}, 5119 (1992); {\bf 46}, 591 (1992).

\item {7.} S. Sachdev, Phys. Rev.
 B {\bf 45}, 389 (1992).

\item {8.} N. Nagaosa and P.A. Lee,
 Phys. Rev. B {\bf 45}, 966 (1992).

\item {9.} Y. Suzumara, Y. Hasegawa,
and H. Fukuyama, J. Phys. Soc. Jpn. {\bf 57}, 2768 (1988).

\item {10.} X.G. Wen, F. Wilczek, and A. Zee, Phys. Rev. B {\bf 39},
11413 (1989).

\item {11.} D.R.T. Jones, J. Kogut,
 D.K. Sinclair, Phys. Rev. D {\bf 19},
1882 (1979).

\item {12.}  M.B. Einhorn and R. Savit,
 Phys, Rev. D {\bf 19}, 1198 (1979).

\item {13.} E. Fradkin and S.H. Shenker, Phys. Rev.
 D {\bf 19}, 3682 (1979).

\item {14.} J.P. Rodriguez, Phys. Rev. B {\bf 49}, 3663 (1994).

\item {15.} V.L. Berezinskii, Zh. Eksp.
Teor. Fiz. {\bf 61}, 1144 (1971)
 [Sov. Phys. JETP {\bf 34}, 610 (1972)].

\item {16.} J.M. Kosterlitz and D.J.
Thouless, J. Phys. C{\bf 6}, 1181 (1973); see also
P. Minnhagen, Rev. Mod. Phys. {\bf 59}, 1001 (1987).

\item {17.} J. Villain, J. Phys. (Paris)
{\bf 36}, 581 (1975).

\item {18.}  J.V. Jos\' e,
L.P. Kadanoff, S. Kirkpatrick and
 D.R. Nelson, Phys. Rev. B {\bf 16},
 1217 (1977).

\item {19.}  C. Itzykson and J.
 Drouffe, {\it Statistical field theory},
vol. 1, chap. 4 (Cambridge Univ.
Press, Cambridge, 1991).

\item {20.} See G. Shirane and R. Birgeneau, in ref. 1.

\item {21.}  J.P. Rodriguez and
Pascal Lederer, Phys. Rev. B {\bf 48}, 16051 (1993).

\item {22.}  J.P. Rodriguez and B. Dou\c cot,
Phys. Rev. B{\bf 45}, 971 (1992).

\item {23.} A.M. Tikofsky, R.B. Laughlin,
and Z. Zou, Phys. Rev. Lett. {\bf 69},
3670 (1992).

\item {24.} P. Nozi\` eres and D. Saint James,
J. Phys. (Paris) {\bf 43}, 1133 (1982).

\item {25.}  Any sum over combinations of sites,
$(r,r^{\prime})$, we define here as $\sum_{(r,r^{\prime})} ... ={1\over
2}\sum_{r,r^{\prime}} ...$ .

\item {26.} S. Trugman, private communications.

\item {27.} A.M. Polyakov, Nucl. Phys. B{\bf 120}, 429 (1977);
{\it Gauge Fields and Strings} (Harwood, New York, 1987).

\item {28.} J.M. Wheatley, T.C. Hsu, and P.W. Anderson,
Phys. Rev. B {\bf 37}, 5897 (1988); S. Chakravarty,
A. Sudbo, P.W. Anderson, and S. Strong, Science {\bf 261}, 337 (1993).

\item {29.} H. Takagi, B. Batlogg, H.L. Kao, J. Kwo, R.J. Cava,
J.J. Krajewski, and W.F. Peck,
Phys. Rev. Lett. {\bf 69}, 2975 (1992).

\vfill\eject
\centerline{\bf Figure Caption}
\vskip 20pt
\item {Fig. 1.}   Shown is the
phase boundary between the
superconducting and the normal phase as a function of
hole carrier concentration, $x$.  The superconducting
 transition temperature is determined by Eq. (9),
with ${\pi\over 2}J_b=(200\,{\rm K})\cdot x/x_0$
and ${\pi\over 2}J_f=(800\,{\rm K})\cdot (1-x/x_0)$,
where $x_0\lsim 1$ denotes the critical concentration
 above which the spinon pairing instability is absent.
The upper set of dashed lines represent cross-over
regions below which ``strange''
metallicity begins to dissappear.

\end